\documentclass[11pt, a4paper]{article}

\usepackage{jcappub}

\def\mr{\mathrm}

\def\eps{\epsilon}
\def\pd{\partial}

\begin{document}

%\begin{flushright}
%{\small YITP-13-20}
%\end{flushright}

\title{Localized Features in Non-Gaussianity from Heavy Physics}
\author{Ryo Saito,$^{1}$} \author{Yu-ichi Takamizu$^{1}$}
\affiliation{
$^1$Yukawa Institute for Theoretical Physics, Kyoto University, \\
~~Kitashirakawa Oiwake-Cho, Sakyo-ku, Kyoto 606-8502, Japan
}

%\begin{abstract}
\abstract{
 We discuss the possibility that we could obtain some hints of the heavy physics during inflation by analyzing local features of the primordial bispectrum. 
 A heavy scalar field could leave large signatures in the primordial spectra through the parametric resonance between its background oscillation and the fluctuations. 
 Since the duration of the heavy-mode oscillations is finite, the effect of the resonance is localized in momentum space.
 In this paper, we show that the bispectrum is amplified when such a resonance occurs, and that the peak amplitude of the feature can be ${\cal O}(10^{1-2})$, or as large as ${\cal O}(10^5)$ depending on the type of interactions.
 In particular, the resonance can give large contributions in finitely squeezed configurations, 
 while the bispectrum cannot be large in the exact squeezed limit. 
 We also find that there is a relation between the scales at which the features appear in the bispectrum and the power spectrum, and that the feature in the bispectrum can be much larger than that in the power spectrum. 
 If correlated features are observed at characteristic scales in the primordial spectra, it will indicate the presence of heavy degrees of freedom. 
 By analyzing these features, we may be able to obtain some information on the physics behind inflation.
 }
 %\end{abstract}

\maketitle

%\tableofcontents
%%%%%%%%%%%%%%%%%%%%%%%%%%%%%%%%%
%Introduction
%%%%%%%%%%%%%%%%%%%%%%%%%%%%%%%%%

\section{Introduction}\label{s:introduction}
Primordial non-Gaussianity is a powerful probe to discriminate inflationary models \cite{Bartolo:2004if}. 
In the simplest setup, i.e. single-field slow-roll inflation models with a canonical kinetic term and Bunch-Davis initial conditions, the primordial fluctuations are predicted to be approximately Gaussian-distributed \cite{Maldacena:2002vr}. 
More generally, for the models of so-called single-clock inflation, where there is only one relevant degree of freedom, it is known that there is a consistency relation relating the bispectrum in the squeezed limit to the tilt of the power spectrum \cite{Creminelli:2004yq, Cheung:2007sv, Creminelli:2011rh, Creminelli:2012ed, Senatore:2012wy, Assassi:2012zq}. 
Thus, a non-Gaussian signal would enable us to narrow down the possible models of inflation. 

To predict the expected non-Gaussian signal from a given model correctly, we need accurate knowledge of the inflationary dynamics as well as the Lagrangian. 
For example, even in the ``single-field" inflation models, where there is a single light scalar field, the consistency relation is not always satisfied. 
It has been argued in Ref. \cite{Namjoo:2012aa} that when the background evolution is provided by a non-attractor solution, the consistency relation is violated while maintaining a scale-invariant power spectrum.
%because the decaying mode plays relevant role in this case. 
To probe the model of inflation, we need to understand well what can happen under a given model.

In standard single-field slow-roll inflation models, non-Gaussian signals in the bispectrum can be classified into three types: local type, equilateral type, and orthogonal type \cite{Cheung:2007st}. 
The recent Planck results have revealed that all these types of non-Gaussianity are consistent with zero and found no deviations from the prediction of the single-field slow-roll inflation models \cite{Ade:2013ydc}.
However, the bispectrum contains  a large number of degrees of freedom and therefore could have much more information.
For example, features localized at a specific scale can be induced in the bispectrum by some temporal events like slow-roll violation or particle production during inflation \cite{Romano:2008rr, Barnaby:2010sq, Leach:2000yw, Saito:2008em, Starobinsky:1992ts, Adams:2001vc, Kaloper:2003nv, Takamizu:2010xy, Ashoorioon:2006wc, Chen:2006xjb, Abolhasani:2012px}.
Localized features can be used to probe the inflationary dynamics.

In general, there could be many scalar degrees of freedom other than the inflaton, even in single-field inflation models.
In a model embedded in supergravity or string theory, for example, such degrees of freedom may appear as moduli fields, Kaluza-Klein modes, or the scalar supersymmetric partner of inflaton. 
Usually, the scalar fields are very heavy, $m \gg H$, so that they are assumed to be stuck in their minima, 
and the model is treated effectively as a single-field model \cite{Yamaguchi:2005qm}. 
Recently, however, it has been pointed out that a heavy scalar field is not necessarily stuck in its potential minimum during inflation \cite{Cremonini:2010ua, Achucarro:2010jv, Achucarro:2010da, Cespedes:2012hu, Achucarro:2012sm, Avgoustidis:2012yc, Chen:2012ge, Pi:2012gf, Burgess:2012dz, Noumi:2012vr, Shiu:2011qw, Gao:2012}. 
It can be displaced from its minimum due to the centrifugal force generated by a turn in the inflaton trajectory.
When the turn is very sharp, even oscillations in the heavy direction can be excited \cite{Shiu:2011qw,Gao:2012}. 
Other possibilities are that oscillations can be excited when the heavy scalar field becomes momentarily light/tachyonic during inflation 
or that they can be excited at the beginning of inflation \cite{Burgess:2002ub}, 
which is natural in the case that inflation occurs after tunneling from a neighboring minimum \cite{Bucher:1994gb, Sasaki:1994yt, Freivogel:2005vv, Yamauchi:2011qq, Sugimura:2011tk, Battefeld:2013xka}, for example.

In the previous paper \cite{Saito:2012pd}, we discussed the possibility that excited oscillation of heavy modes can leave non-negligible signatures in the power spectrum through derivative couplings, without spoiling inflation (see also Refs. \cite{Chen} for models with only gravitational couplings). 
We saw that the primordial fluctuations can be enhanced deep in the horizon, $k/a \sim m \gg H$, by the parametric resonance with the excited oscillations. 
In this paper, we estimate the resonant feature in the bispectrum within the same setup.
This gives an example where resonant non-Gaussianity \cite{Chen:2008wn, Flauger:2010ja, Chen:2010bka, Behbahani:2011it, Gwyn:2012pb} is produced in a realistic model. 
Since the duration of heavy-mode oscillations is finite, the effect of the resonance is localized in momentum space.  
%\footnote{See e.g. Refs. \cite{Romano:2008rr, Barnaby:2010sq, Leach:2000yw, Saito:2008em, Starobinsky:1992ts, Adams:2001vc, Kaloper:2003nv, Ashoorioon:2006wc, Chen:2006xjb, Abolhasani:2012px} for possible other effects leaving localized features in the primordial spectra.}
We will show that a large feature could be induced in the bispectrum even when the feature in the power spectrum is too small to be observed. 
We also investigate the behavior of the bispectrum in the squeezed limit. 
As in the case that the consistency relation is satisfied, 
it can be shown that the bispectrum in the squeezed limit cannot be large unless the modification to the power spectrum is large. 
However, we will also show that, for finitely squeezed configurations,  the bispectrum can be greatly enhanced in comparison to the usual single-field inflation model (a model with a single light scalar field).
Since we cannot observe the exact squeezed limit in actual observations,  the enhancement could be practically important, as recently pointed out in Ref. \cite{Flauger:2013hra}, where the resonant non-Gaussianity from a modulated potential is considered as a concrete example. 
The detection of such features would therefore indicates the presence of heavy degrees of freedom during inflation, and by analyzing them we may hope to improve our understanding of the physics behind inflation.

The organization of this paper is as follows.
In \S\ref{s:Model}, we briefly review the model presented in Ref. \cite{Saito:2012pd} to realize an efficient enhancement of the fluctuations in the inflaton field and estimate the feature induced in the power spectrum. In \S\ref{s:FB}, we discuss the resonant enhancement of the bispectrum. Finally, we provide a summary of this paper in \S\ref{s:summary}.

%%%%%%%%%%%%%%%%%%%%%%%
\section{Overview of the Model}\label{s:Model}
%%%%%%%%%%%%%%%%%%%%%%%

%%%%%%%%%%%%%%%%%%%%%%%
	\subsection{The Model}
 First, we briefly review the model presented in Ref. \cite{Saito:2012pd}.  We consider a model with a heavy scalar field with mass $m \gg H$, $\chi (\equiv \phi^{(2)})$, which derivatively couples to the inflaton field, $\phi(\equiv \phi^{(1)})$, as
	\begin{align}
		S_{m} &\equiv -\int \mr{d}x^4 \sqrt{-g}P(X^{IJ},\phi^K), \qquad (I,J,K=1,2) \label{eq:generalaction}\\
		&= -\int \mr{d}x^4 \sqrt{-g}\left[ \frac{1}{2}(\pd \phi)^2 + V(\phi) + \frac{1}{2}(\pd \chi)^2 + \frac{m^2}{2}\chi^2 + K_n + K_d + {\cal O}\left(X\phi^2/\Lambda_n^2, X^3/\Lambda_d^8 \right) \right], \label{eq:action}
	\end{align}
with $X^{IJ} \equiv -\pd_{\mu} \phi^I\pd^{\mu} \phi^J/2$. 
Here, we have assumed that the derivative couplings at the leading order are given by,
	\begin{align}
		K_n  &\equiv \frac{\lambda_n}{2\Lambda_n}\chi(\pd \phi)^2, \label{eq:ncouple}
	\end{align}
and
	\begin{align}
		K_d  &\equiv \frac{\lambda_{d1}}{4\Lambda_d^4}(\pd \chi)^2(\pd \phi)^2 + \frac{\lambda_{d2}}{4\Lambda_d^4}(\pd \chi \cdot \pd \phi)^2, \label{eq:dcouple}
	\end{align}
with the dimensional parameters $\Lambda_n, \Lambda_d$ and the dimensionless parameters $\lambda_n, \lambda_{di}~(i=1,2)$. 
 The derivative couplings in Eq. (\ref{eq:action}) generally appear in the action from the effective-field-thoery point of view \cite{Weinberg:2008hq, Khosravi:2012qg} . 
The derivative couplings $K_n$ and $K_d$ provide the most general couplings between the inflaton field and the heavy scalar field at the leading order in $1/\Lambda_n$ and $1/\Lambda_d$ in a model with the parity symmetry, $\phi \to -\phi$, and the shift symmetry, $\phi \to \phi+c$. 
 The approximate shift symmetry is usually assumed to ensure the flatness of the inflaton potential. 
 In addition, we have assumed the parity symmetry to forbid the kinetic mixing, which obscures the difference between the light and heavy fields. 
 Thus, we first focus on the couplings $K_n$ and $K_d$, though brief comments are presented in \S\ref{ss:SVI} on expected features in the bispectrum when a symmetry-violating interaction becomes relevant. In general, higher-order terms in $\Lambda_n$ and $\Lambda_d$ are also expected to appear. 
 To ensure that contributions from these terms can be treated perturbatively, we assume here that the background fields satisfy the following conditions,
 	\begin{equation}\label{eq:irrelevant}
		\chi \ll \Lambda_n, \quad \dot{\phi},~\dot{\chi} \ll \Lambda_d^2.
	\end{equation}
We have also suppressed the terms $(\pd \phi)^4$ and $(\pd \chi)^4$ in Eq. (\ref{eq:action}), because they have little effects on our analysis for the background evolution and the power spectrum. 
The higher-order terms and the self interaction $(\pd \phi)^4$ can be important for the bispectrum. 
Features in the bispectrum induced by these interactions are discussed in \S\ref{ss:SVI}. 
We can also find many specific models which have the couplings $K_n$ and $K_d$, such as DBI inflation \cite{Kobayashi:2012kc}. 
In these cases, the conditions (\ref{eq:irrelevant}) are not mandatory.

Since $\chi$ can generally decay into other particles, we assume that $\chi$ decays with a rate $\Gamma$, which satisfies $H \ll \Gamma \ll m$. 
To ensure the slow-roll inflation, we further make two assumptions. 
First, the inflaton potential is sufficiently flat,
	\begin{align}\label{eq:slowroll}
		\eps_V \ll 1,~|\eta_V| \ll 1,
	\end{align}
where
	\begin{equation}
		\eps_V \equiv \frac{M_p^2}{2}\left(\frac{V'}{V}\right)^2, \quad \eta_V \equiv M_p^2\frac{V''}{V},
	\end{equation}
are the slow-roll parameters. Here, $M_p=2.4 \times 10^{18}~\mr{GeV}$ is the reduced Planck mass. 
Second, the heavy scalar field $\chi$ is subdominant,
	\begin{align}\label{eq:sub}
		f_{\chi} \ll 1,
	\end{align}
where
	\begin{equation}\label{eq:fchi}
		f_{\chi} \equiv \frac{\rho_{\chi}}{\rho} \simeq \frac{\dot{\chi}^2+m^2\chi^2}{6M_p^2H^2},
	\end{equation}
is the fraction of its energy density to the total one.

\begin{figure}[t]
		\vspace*{0.2cm}
		\centering
		\includegraphics[width=.55\linewidth]{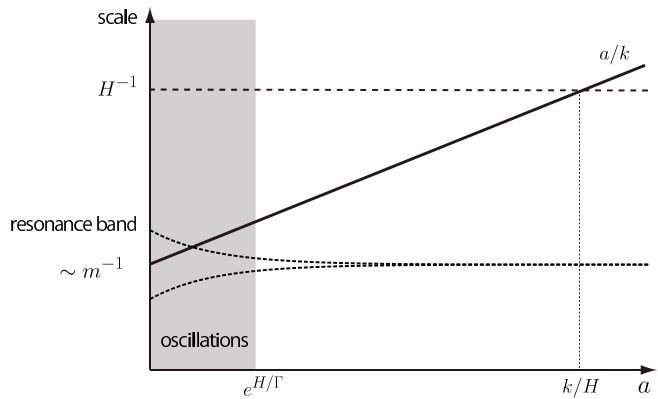}
		\caption{The fluctuations in the inflaton field can be amplified through the parametric resonance with the background oscillation of the heavy scalar field (\ref{eq:hbgsol}). As depicted by the thick line, a mode is redshifted by the cosmic expansion. Since only the modes that have crossed the resonance band during the oscillation are amplified, the features in the primordial spectra are localized in momentum space.}
		\label{fig:evolution}
	\end{figure}

Provided that the conditions (\ref{eq:irrelevant}), (\ref{eq:slowroll}), and (\ref{eq:sub}) are satisfied, the background evolution of the inflation field is given by the slow-roll solution,
	\begin{align}
		\pi_{\phi}(t) \simeq -\frac{V'}{3H}, \label{eq:ibgsol}
	\end{align}
where $\pi_{\phi}$ is the conjugate momentum for the inflaton field,
	\begin{align}
		\pi_{\phi} \equiv \hat{z}_{\phi}^2\dot{\phi}, \label{eq:icm}
	\end{align}
with
	\begin{align}
		z_{\phi}^2 \equiv a^3\left[ 1 + \lambda_n\frac{\chi}{\Lambda_n} + \left(\lambda_{d1}+\lambda_{d2}\right)\frac{\dot{\chi}^2}{2\Lambda_d^4} \right],
	\end{align}
and $\hat{z}_{\phi}^2 \equiv z_{\phi}^2/a^3$. 
The heavy scalar field oscillates with a frequency of the mass scale $m$,
	\begin{align}
		\chi(t) \simeq \chi_0 e^{-\Gamma t}\cos(mt)\theta(t), \label{eq:hbgsol}
	\end{align}
with $\theta(t)$ being the Heaviside function, where we have assumed that the oscillation is excited instantaneously at $t=0$.
The fluctuations are enhanced through a resonance with the background oscillation of the heavy scalar field (\ref{eq:hbgsol}).
We showed the basic picture of the resonant enhancement in Fig. \ref{fig:evolution}.

%%%%%%%%%%%%%
%%%%%%%%%%%%%

 For a later convenience, we introduce parameters which represent the magnitude of the derivative couplings,
 	\begin{align}
		q_n &\equiv \lambda_n\frac{\chi_0}{\Lambda_n}, \\
		q_{d1} &\equiv \lambda_{d1}\frac{m^2\chi_0^2}{2\Lambda_d^4}, \quad q_{d2} \equiv \lambda_{d2}\frac{m^2\chi^2_0}{2\Lambda_d^4},
	\end{align}
which are related to the $q$-parameters introduced in Ref. \cite{Saito:2012pd}.
We will also use $q$ to represent the order of the $q$-parameters above. 
Note that the $q$-parameters are less than unity under the conditions (\ref{eq:irrelevant}).
They can be expressed in terms of the energy fraction of the heavy scalar field, $f_{\chi}$, as
	\begin{align}
		q_{n} \simeq \sqrt{2}\lambda_{n}\left(\frac{E_{\rm inf}}{m}\right)\left(\frac{E_{\rm inf}}{\Lambda_n}\right)f_{\chi, t=0}^{\frac{1}{2}},
	\end{align}
for the couplings $K_n$ and
	\begin{align}
		q_{d1} \simeq \lambda_{d1}\left(\frac{E_{\rm inf}}{\Lambda_d}\right)^4f_{\chi, t=0}, \quad q_{d2} \simeq  \lambda_{d2}\left(\frac{E_{\rm inf}}{\Lambda_d}\right)^4f_{\chi, t=0},
	\end{align}
for the couplings $K_d$ , where $E_{\rm inf}$ is the energy scale of the inflation, $E_{\rm inf} \equiv (3M_p^2H^2)^{\frac{1}{4}} \simeq V^{\frac{1}{4}}$.

%%%%%%%%%%%%%%%%%%%%%%% 
	\subsection{Power Spectrum}\label{ss:power}
 Before investigating the features in the bispectrum, we first estimate the effect of the resonance on the power spectrum. 
 To obtain an analytic expression, here, we use the perturbative method assuming that the $q$-parameters are sufficiently small.
 
 To eliminate the gauge degrees of freedom in the metric fluctuations, we use the flat gauge, where the spatial metric becomes $a^2\delta_{ij}$. 
 In this gauge, we can make some simplifications in estimating the power spectrum. 
 First, the metric fluctuations can be neglected because the resonance takes place deep in the horizon. 
 Moreover, we can also neglect the fluctuations in the heavy scalar field, $\delta \chi$. 
 This is because $\delta \chi$ oscillates with a frequency $\omega>m$ and therefore does not satisfy the resonance condition. 
In addition, the mass term suppresses the amplitude of $\delta \chi$ at $k/a \sim m$ compared with that of the inflaton field.
 
 Neglecting the fluctuations in the metric and the heavy scalar field, the second-order action for the fluctuations in the inflaton field, $\varphi$, can be written as,
 	\begin{align}
		S_2 \simeq \int \mr{d}t\mr{d}^3x~ \frac{z_{\phi}^2}{2}\left[ \dot{\varphi}^2 - c_s^2(\nabla \varphi)^2/a^2 \right], \label{eq:2action}
	\end{align}
where
	\begin{align}
		c_s^2 = 1 - \frac{\lambda_{d2}\dot{\chi}^2}{2\Lambda_d^4} + O\left(\frac{\dot{\chi}^4}{\Lambda_d^8}\right),
	\end{align}
is the speed of sound for the fluctuations in the inflaton field. 
Note that we can neglect terms like $\dot{\phi}^2/\Lambda_d^4$ in the action (\ref{eq:2action}), 
which could be induced by the self interaction $(\pd \phi)^4/\Lambda_d^4$, 
because they are higher order in the $q$-parameters.

 Introducing the canonically normalized variable $v \equiv z_{\phi}\varphi$, the action (\ref{eq:2action}) can be rewritten as,
 	\begin{align}
		S_2 = \int \mr{d}t\mr{d}^3x~ \frac{1}{2}\left[ \dot{v}^2 - \left(\frac{c_s^2\nabla^2}{a^2} - \frac{\ddot{z}_{\phi}}{z_{\phi}} \right)v^2 \right] , \label{eq:2actionc}
	\end{align}
and then the full Hamiltonian as, 
	\begin{align}
		H^{(2)} = \int \mr{d}^3x~ \frac{1}{2}\left[ \dot{v}^2 + \left(\frac{c_s^2\nabla^2}{a^2} - \frac{\ddot{z}_{\phi}}{z_{\phi}} \right)v^2 \right] . \label{eq:2hamiltonian}
	\end{align}
Substituting the background solution (\ref{eq:hbgsol}), the quantities $c_s^2$ and $\ddot{z}_{\phi}/z_{\phi}$ can be estimated as,
	\begin{align}
		c_s^2 &\simeq 1 + \frac{q_{d2}}{2}e^{-2\Gamma t}\cos(2mt), \label{eq:sonic}\\
		\frac{\ddot{z}_{\phi}}{z_{\phi}} &\simeq m^2\left[q_n e^{-\Gamma t}\cos(mt) + (q_{d1}+q_{d2})e^{-2\Gamma t}\cos(2mt) + O\left(\frac{H}{m}\right)\right]. \label{eq:efmass}
	\end{align}
In the following analysis, we neglect the ${\cal O}(H/m)$ terms in Eq. (\ref{eq:efmass}) assuming that $q$ is larger than $H/m$.
 Taking the free Hamiltonian as the terms independent of the derivative couplings in Eq. (\ref{eq:2hamiltonian}),
the interaction Hamiltonian is given by, 
\footnote{Note that infinitely many terms would appear in the interaction Hamiltonian if we define the free Hamiltonian as terms independent of the derivative couplings without introducing the canonically normalized variable $v$.}

	\begin{align}\label{eq:intH2}
		H_{I}^{(2)} \equiv \frac{m^2}{2(2\pi)^3}\int \mr{d}^3k~ C_k^{(2)}v_{{\bf k}}v_{\bf -k},
	\end{align}
where we have moved to Fourier space,
	\begin{align}
		v(x) = \frac{1}{(2\pi)^3}\int{\rm d}^3k~v_{\bf k}e^{i{\bf k}\cdot{\bf x}}.
	\end{align} 
Here, the coefficient $C_k^{(2)}$ is given in terms of the $q$-parameters as,
	\begin{align}\label{eq:Ck2}
		C_k^{(2)} \equiv C_k^{(2;n)}e^{-\Gamma t}\cos(mt) + C^{(2;d)}e^{-2\Gamma t}\cos(2mt),
	\end{align}
where
	\begin{align}
		C_k^{(2;n)} &\equiv -q_n, \label{eq:c2n}\\
		C_k^{(2;d)} &\equiv q_{d2}\left[\frac{1}{2}\left(\frac{k}{am}\right)^2 - 1\right] - q_{d1}. \label{eq:c2d}
	\end{align}
Hence, using the in-in formalism, the correction to the power spectrum can be estimated as, 
	\begin{align}
		\Delta \langle \varphi_{\bf k}(t)\varphi_{\bf k'}(t) \rangle &\simeq i\int_0^t {\rm d}t'~\langle [H_{I}^{(2)}(t'), \varphi_{\bf k}(t)\varphi_{\bf k'}(t)] \rangle,
	\end{align}
at the leading order in the $q$-parameters, where the fields in LHS are operators in the Heisenberg picture while, in RHS, they represent operators in the interaction picture. 
For brevity, we denote them in the same way 
since it will be clear which they represent from the context.

In terms of the dimensionless power spectrum, ${\cal P}_{\varphi}$, defined through,
	\begin{align}\label{eq:defpower}
		\langle \varphi_{{\bf k}}(t)\varphi_{{\bf k}'}(t) \rangle \equiv (2\pi)^3\delta^3({\bf k+k'})\frac{2\pi^2{\cal P}_{\varphi}}{k^3},
	\end{align}
the correction is expressed as,
	\begin{align}\label{eq:deltaPphi}
		\Delta {\cal P}_{\varphi}(k)/{\cal P}_{\varphi} =  I_{k}^{(2)},
	\end{align}
where
	\begin{align}\label{eq:I2}
		I_{k}^{(2)} \equiv -2m^2{\rm Re}\left[i\int_0^t {\rm d}t'~ C_k^{(2)}u_k(t')^2\right].
	\end{align}
Here, the estimation time $t$ should be taken sufficiently after the decay of the oscillation. 
The function $u_k$ is the mode function for $v_{{\bf k}}$, which is defined through,
	\begin{align}
		v_{\bf k}(t) = u_k(t)a_{{\bf k}} + u_k^{\ast}(t){a_{-{\bf k}}}^{\dagger},
	\end{align}
for the creation/annihilation operators.
Though the state could be excited depending on the excitation mechanism of the oscillation, we simply assume that the state was in the vacuum state at the excitation of the oscillation. 
Even if the state was excited, 
the resonance is not spoiled except for some specific excited states. 
Assuming that the state is not excited, the mode function is provided by,
	\begin{align}
		u_k(t) = \sqrt{\frac{a}{2k}}\left(1-\frac{i}{k\tau}\right)e^{-ik\tau},
	\end{align}
where $\tau$ is the conformal time. 
Because the resonance occurs deep in the horizon, we can approximate the mode functions in the integral (\ref{eq:I2}) as,
	\begin{align}\label{eq:MFsub}
		u_k(t) \simeq \sqrt{\frac{a}{2k}}e^{-ik\tau}.
	\end{align}
Hence, $I_k^{(2)}$ can be approximated as,
	\begin{align}\label{eq:I2app}
		I_{k}^{(2)} \simeq -\int_0^z {\rm d}z' \left(\frac{k}{am}\right)^{-1}C_k^{(2)}\sin(2k\tau'), \quad (z \equiv mt).
	\end{align}
The resonance occurs between the factor $\sin(2k\tau')$ and the oscillatory components, $\cos(mt')$ and $\cos(2mt')$, in $C_k^{(2)}$ (see Eq. (\ref{eq:Ck2})). 
If the negative energy mode exists, it would introduce an additional oscillatory component in Eq. (\ref{eq:I2app}) with a constant phase shift. 
However, this additional component does not spoil the resonance 
unless the amplitude of the negative energy mode and the phase shift have specific values for the scales of the resonance.

Here, we evaluate $I^{(2)}_k$ for the couplings $K_n$. 
Replacing the parameters $(q_n,m/2,\Gamma/2)$ by $(q_d,m,\Gamma)$, we can obtain a similar result for the couplings $K_d$. 
Using the product-to-sum identities for the trigonometric functions, we have two oscillatory functions in Eq. (\ref{eq:I2app}).
The integration has a resonant contribution from one of the two in the interval around the stationary point of the phase function, $\theta^{(2;n)} \equiv mt - 2k\tau$. 
Taking the derivative of the phase function $\theta^{(2;n)}$, we can find that the resonance occurs at time $t_{\ast}$ where,
	\begin{align}
		\frac{k}{a(t_{\ast})} = \frac{m}{2}.
	\end{align}
Hence, the resonance occurs $\Delta N \equiv \ln(m/2H)$ e-folds before the horizon crossing of the modes.
%for the $K_n$ couplings and
%	\begin{align}
%		\frac{k}{a(t_{\ast})} = m,
%	\end{align}
%for the $K_d$ couplings. 
Solving the above equation, the value of $z$ at the resonance is estimated to be
	\begin{align}
		z_{\ast} \simeq \frac{m}{H}\ln\left(\frac{2k}{a_0m}\right),
	\end{align}
%for the $K_n$ couplings and 
%	\begin{align}
%		z_{\ast} \simeq \frac{m}{H}\ln\left(\frac{k}{a_0m}\right),
%	\end{align}
%for the $K_d$ couplings, respectively. 
where $a_0$ is the scale factor at the onset of the oscillation. 
Then, the integral can be evaluated as,
\footnote{Note that we have used $2k$ for the argument of $I^{(n)}_k$ for later convenience. The factor $2k$ in Eq. (\ref{eq:Inorm}) corresponds to $k+k'$ if we don't take into account the delta function in Eq. (\ref{eq:defpower}). Similarly, the scaling function $I^{(n)}_{K}$ with $K=k_1+k_2+k_3$ appears for the bispectrum.}
	\begin{align}
		I_{k}^{(2)} \simeq C_{k,\ast}^{(2;n)}{\rm Im} I^{(n)}_{2k},
	\end{align}
where
	\begin{align}\label{eq:Inorm}
		I^{(n)}_{2k} \equiv \int_0^z {\rm d}z' e^{-\frac{\Gamma}{m}z' + i(z'-2k\tau')},
	\end{align}
and $C_{k, \ast}^{(2;n)}$ is the coefficient function (\ref{eq:c2n}) at the resonance, $t = t_{\ast}$.
The duration of the resonance can be roughly estimated by $\Delta z \equiv m/\sqrt{\ddot{\theta}}$, which is given by
	\begin{align}
		\Delta z^{(2;n)} &= \sqrt{m/H}, \label{eq:durationN}
%		\Delta z^{(2;d)} &= \sqrt{2m/H}. \label{eq:durationD} 
	\end{align}
for the couplings $K_n$.
Approximating the oscillating factor $e^{i(z'-2k\tau')}$ in Eq. (\ref{eq:Inorm}) by a top-hat function, then, we can roughly estimate the contribution to $I^{(n)}_{2k}$ from the resonance through the couplings $K_n$ as,
	\begin{align}
		I_{2k}^{(n)} &\simeq C_{k,\ast}^{(2;n)}\sin\theta^{(2;n)}_{\ast}\int_{\max(0,z_{\ast}-\sqrt{\frac{m}{H}})}^{\max(0,z_{\ast}+\sqrt{\frac{m}{H}})} {\rm d}z' ~e^{-\frac{\Gamma}{m} z'} \\
		&\equiv C_{k,\ast}^{(2;n)}\sin\theta^{(2;n)}_{\ast}I_{2k, \text{approx}}^{(n)} \\
		&=
		{\displaystyle
		\begin{cases}
			0 & \left(\ln\left(\frac{2k}{a_0m}\right) < -\sqrt{\frac{H}{m}}\right), \\
			-\frac{m}{\Gamma}\left[1 - \left(\frac{2k}{a_0m}\right)^{-\frac{\Gamma}{H}}e^{-\frac{\Gamma}{\sqrt{Hm}}} \right]\sin\theta^{(2;n)}_{\ast}q_n & \left( -\sqrt{\frac{H}{m}} < \ln\left(\frac{2k}{a_0m}\right) < \sqrt{\frac{H}{m}}\right), \\
			-\frac{2m}{\Gamma}\left(\frac{2k}{a_0m}\right)^{-\frac{\Gamma}{H}}\sinh\left(\frac{\Gamma}{\sqrt{Hm}}\right)\sin\theta^{(2;n)}_{\ast}q_n & \left(\ln\left(\frac{2k}{a_0m}\right) > \sqrt{\frac{H}{m}}\right),
		\end{cases}
		} \label{eq:Inormapprox}
		\end{align}
where $\theta^{(2;n)}_{\ast}$ is the phase function at the resonance, 
	\begin{align}
		\theta^{(2;n)}_{\ast} \simeq \frac{m}{H}\left[\ln\left(\frac{2k}{a_0m}\right)+1\right].		
	\end{align}
Hence, the power spectrum has a peak at $k_p/a_0 \simeq me^{\sqrt{H/m}}/2$ with an amplitude,
	\begin{align}\label{eq:correction2}
		\Delta {\cal P}_{\varphi}^{(n)}(k_p)/{\cal P}_{\varphi} \sim \frac{m}{\Gamma}\left(1 - e^{-\frac{2\Gamma}{\sqrt{Hm}}}\right)q_n.
	\end{align}
In particular, when the decay rate is small, $\Gamma \ll \sqrt{Hm}$, the peak amplitude becomes,
	\begin{align}\label{eq:correction2lim}
		\Delta {\cal P}_{\varphi}^{(n)}(k_p)/{\cal P}_{\varphi} \sim 2q_n\sqrt{\frac{m}{H}},
	\end{align}
which is of the order of $2q_n\Delta z^{(2;n)}$. 

\begin{figure}[t]
		\centering
		\includegraphics[width=0.6\linewidth]{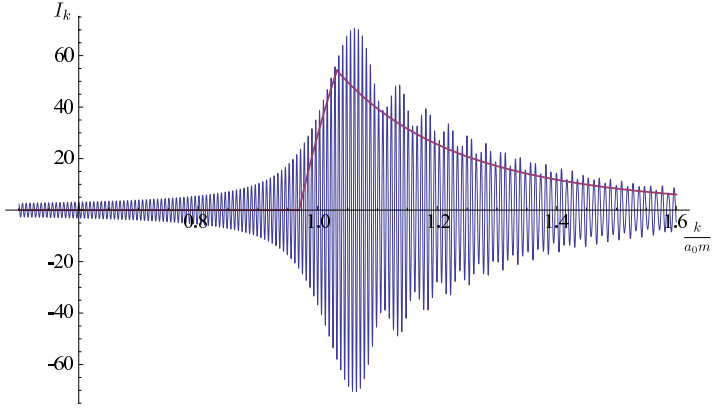}
		\caption{
		The scaling function $I_k^{(n)}$ (Eq. (\ref{eq:Inorm})), which determines the scale dependence of the resonant enhancement. 
		The thick line represents the approximate function, $I_{k,\text{approx}}^{(n)}$. The function has a peak at $k/(a_0m) \simeq e^{\sqrt{H/m}}$.
		In the plot, we have set $\Gamma = 5H$ and $m = 10^3 H$.
		}
		\label{fig:power}
	\end{figure}

 In Fig. \ref{fig:power}, we showed the scaling function $I_k^{(n)}$ (Eq. (\ref{eq:Inorm})). 
 The function $I^{(n)}_k$ has a peak at $k/a_0m \sim 1$.
As can be seen in the figure, the approximate function $I_{k,\text{approx}}^{(n)}$ introduced in Eq. (\ref{eq:Inormapprox}) gives a reasonable fit of $I_k^{(n)}$.

%%%%%%%%%%%%%%%%%%%%%%%%%%%%%%%
%%%%%%%%%%%%%%%%%%%%%%%%%%%%%%%

 Once the correlation function for the inflaton field are obtained, we can calculate those for the comoving curvature fluctuations $\zeta$ by using the gauge transformation. 
\footnote{Note that the notation ${\cal R}$ is also used to denote the comoving curvature fluctuations. 
The variable $\zeta$ here corresponds to ${\cal R}$ in Ref. \cite{Bassett:2005xm}, for example.}
 After the oscillation has damped out, the gauge transformation is given as usual,
 	\begin{align}\label{eq:gt}
		\zeta = \frac{H}{\dot{\phi}}\varphi,
	\end{align}
at the linear order. 
At the non-linear level, in general, we can obtain the correlation functions of $\zeta$ by using $\delta N$-formalism \cite{Sasaki:1995aw, Wands:2000dp, Lyth:2004gb, Tanaka:2010km}
from those of $\varphi$ on superhorizon scales. 
%where the oscillation damped out. 
Since $\zeta$ is proportional to $\varphi$ at the linear order, the correction to the power spectrum for $\zeta$, $\Delta {\cal P}_{\zeta}/{\cal P}_{\zeta}$, is provided also by Eq. (\ref{eq:deltaPphi}):
	\begin{align}\label{eq:deltaPzeta}
		\Delta {\cal P}_{\zeta}(k)/{\cal P}_{\zeta} =  I_{k}^{(2)} \simeq C_{k,\ast}^{(2;n)}\sin\theta^{(2;n)}_{\ast}I_{2k, \text{approx}}^{(n)}.
	\end{align}
Hence, the peak amplitude is given by Eq. (\ref{eq:correction2}) (or Eq. (\ref{eq:correction2lim})):
 	\begin{align}
		\Delta {\cal P}_{\zeta}^{(n)}(k_p)/{\cal P}_{\zeta} &\sim \frac{m}{\Gamma}\left(1 - e^{-\frac{2\Gamma}{\sqrt{Hm}}}\right)q_n \label{eq:correction2zeta} \\
		&\sim 2q_n\sqrt{\frac{m}{H}} \quad (\text{for } \Gamma \ll \sqrt{Hm}).
	\end{align}

 We close this section with a comment on the perturbativity. 
 Even when the $q$-parameters are smaller than unity, 
 the higher-order terms in the expansion with respect to the interactions can be comparable to the leading term 
 when the duration of the resonance $\Delta t$ is sufficiently long. 
 This is because the resonance coherently accumulates the effects of the interaction.
% This is because the interaction Hamiltonian appears with time integration. 
 Then, the order parameter of the expansion is given by $H_I^{(2)}\Delta t$, 
 which could be large even when $q$ is very small.
 The perturbative method is valid only when the correction to the power spectrum (\ref{eq:correction2lim}) is less than unity. 
 When this condition is not satisfied, we should solve the full equation of motion numerically to get the mode function as done in Ref. \cite{Saito:2012pd}.
 
 Other than the resonance, the derivative couplings could induce the loop corrections to the power spectrum and the Green function. 
 To ensure that the corrections are suppressed, the model (\ref{eq:action}) should have a cutoff at the energy scale $\Lambda_{\rm cut} \equiv \min(2\pi\Lambda_n, (2\pi)^{1/2}\Lambda_d)$. 
  \footnote{Here, we have assumed that the coupling constants of the derivative couplings, $\lambda$, are of the order of unity.}
 Hence, to perform the calculation perturbatively at the resonance scale, $k/a \sim m$, the mass scale should not exceed the cutoff scale, $m < \Lambda_{\rm cut}$.

%%%%%%%%%%%%%%%%%%%%%%
\section{Features in the Bispectrum}\label{s:FB}
%%%%%%%%%%%%%%%%%%%%%%

 Next, we investigate the effect of the resonance on the bispectrum,
 	\begin{align}\label{eq:dlB}
		\langle \zeta_{{\bf k}_1}(t)\zeta_{{\bf k}_2}(t)\zeta_{{\bf k}_3}(t) \rangle \equiv (2\pi)^7\delta^3({\bf k_1+k_2+k_3}){\cal P}_{\zeta}^2\frac{{\cal B}_{\zeta}}{k_1^2k_2^2k_3^2}.
	\end{align}
After showing which cubic interactions are important for the resonance, we estimate the correction to the bispectrum perturbatively by using the in-in formalism.

 We use the dimensionless quantity ${\cal B}_{\zeta}$ to represent the amplitude of the bispectrum, which is made dimensionless by the factor $k_1^2k_2^2k_3^2$ as in Ref. \cite{Babich:2004gb}. 
 Note that this normalization factor differs from that in the definition of the local-type $f_{NL}$. 
 In addition, we use the uncorrected power spectrum ${\cal P}_{\zeta}$ to define the amplitude ${\cal B}_{\zeta}$. 
% \footnote{Actually, it wouldn't matter if we use the corrected power spectrum as a normalization factor because the difference is higher order in the $q$-parameters.}

%%%%%%%%%%%%%%%%%%%%%% 
	\subsection{Interactions Relevant to the Resonance}
 Here, we discuss which cubic interactions play important roles in the resonance. 
 First, they should obviously contain oscillatory components. 
 Secondly, the interactions with more derivatives become important for the resonance. 
 This is because the resonance occurs deep in the horizon, $k/a \sim m \gg H$, where the fluctuations oscillate with high frequencies. 
 Note that the fluctuations in the heavy scalar field can be neglected at the leading order as in the case of the power spectrum. 
 Hence, we do not need to consider the mixing between the inflaton field and the heavy scalar field.
 
 Taking into account the above points, we can pick up the relevant interactions as (see Appendix \ref{s:RI}),
 	\begin{align}\label{eq:intHphi}
		H_I^{(3)} &=  \frac{a^3}{M_p}\sqrt{\frac{\epsilon_V}{2}}\int {\rm d}^3x~\biggl\{ -\frac{\hat{z}_{\phi}^2}{2}\left[  (3-2c_s^2)\dot{\varphi}^2 + \frac{c_s^2}{a^2}(\nabla \varphi)^2 \right]\varphi \nonumber \\
		&\qquad +  \left[\hat{z}_{\phi}^2 + 2(q_{d1}+q_{d2})e^{-2\Gamma t}\sin^2(mt)\right]\nabla^i(\nabla^{-2}\dot\varphi)\nabla_i\varphi\dot{\varphi}  \biggr\}.
	\end{align}
Note that we need to include the gravitational couplings to obtain the interactions (\ref{eq:intHphi}) 
because the couplings $K_n$ and $K_d$ do not contain cubic interactions of $\phi$ such as $(\pd \chi \cdot \pd \phi)(\pd \phi)^2$ due to the parity symmetry. 
This is why the interactions (\ref{eq:intHphi}) are suppressed by the Planck scale, $M_p$. 
However, direct cubic interactions exist if the self interaction $(\pd \phi)^4$ or higher-order interactions are introduced, 
though we have neglected them in the previous section 
because they have little effects on the background evolution and the power spectrum.
They are also expected to arise if we relax the requirement of the parity symmetry.
We will discuss these possibilities in \S\ref{ss:SVI}.

In terms of the canonically normalized variable $v \equiv z_{\phi}\varphi$, the interaction Hamiltonian can be written as,
	\begin{align}\label{eq:intHv}
		H_I^{(3)} &= \frac{1}{a^{\frac{3}{2}}M_p(2\pi)^6}\sqrt{\frac{\epsilon_{V}}{2}}\int {\rm d}^3k_1{\rm d}^3k_2{\rm d}^3k_3\delta^3({\bf k_1+k_2+k_3}) \left[m^2C^{(3O)}_{\bf k_1 k_2 k_3}v_{\bf k_1}v_{\bf k_2}v_{\bf k_3} \right. \nonumber \\
		&\qquad \left. + mC^{(3I)}_{\bf k_1 k_2 k_3}v_{\bf k_1} \left( v_{\bf k_2}v_{\bf k_3}\right)^{\cdot} + C^{(3II)}_{\bf k_1 k_2 k_3}v_{\bf k_1}\dot{v}_{\bf k_2}\dot{v}_{\bf k_3} \right].
	\end{align}
Here, each coefficients are decomposed into the $K_n$- and $K_d$-coupling parts as,
	\begin{align}\label{eq:C3koii}
		C^{(3i)}_{\bf k_1 k_2 k_3} &\equiv C^{(3i;n)}_{\bf k_1 k_2 k_3}e^{-\Gamma t}\cos(mt) + C^{(3i;d)}_{\bf k_1 k_2 k_3} e^{-2\Gamma t}\cos(2mt) , \qquad (i=O,II),
	\end{align}
and
	\begin{align}\label{eq:C3ki}
		C^{(3I)}_{\bf k_1 k_2 k_3} &\equiv C^{(3I;n)}_{\bf k_1 k_2 k_3}e^{-\Gamma t}\sin(mt) + C^{(3I;d)}_{\bf k_1 k_2 k_3} e^{-2\Gamma t}\sin(2mt),
	\end{align}
where
	\begin{align}
		C^{(3O;n)}_{\bf k_1 k_2 k_3} &\equiv -\frac{q_n}{4}\frac{{\bf k_1 \cdot k_2}}{(am)^2}, \\
		C^{(3I;n)}_{\bf k_1 k_2 k_3} &\equiv -\frac{q_n}{2}\left(\frac{1}{2} - \frac{{\bf k_1 \cdot k_2}}{k_2^2}\right), \\
		C^{(3II;n)}_{\bf k_1 k_2 k_3} &\equiv \frac{q_n}{2}\left(\frac{1}{2} - \frac{{\bf k_1 \cdot k_2}}{k_2^2}\right),
	\end{align}
and
	\begin{align}
		C^{(3O;d)}_{\bf k_1 k_2 k_3} &\equiv \frac{q_{d1}+3q_{d2}}{8}\frac{{\bf k_1 \cdot k_2}}{(am)^2}, \\
		C^{(3I;d)}_{\bf k_1 k_2 k_3} &\equiv -\frac{q_{d1}+q_{d2}}{2}\left(\frac{1}{2} - \frac{{\bf k_1 \cdot k_2}}{k_2^2}\right), \\
		C^{(3II;d)}_{\bf k_1 k_2 k_3} &\equiv -\frac{q_{d1}}{4}\left(\frac{1}{2} + \frac{3{\bf k_1 \cdot k_2}}{k_2^2}\right) + \frac{3q_{d2}}{4}\left(\frac{1}{2} - \frac{{\bf k_1 \cdot k_2}}{k_2^2}\right).
	\end{align}
Here, we have extracted only oscillatory components from the interactions (\ref{eq:intHphi}) and neglected the higher-order terms in the $q$-parameters and $H/m$.

%%%%%%%%%%%%%%%%%%%%%%%%%%%%%%%%

	\subsection{Resonant Enhancement of the Bispectrum}
 Provided the interaction Hamiltonian (\ref{eq:intHv}), we can estimate the correction to the bispectrum by using the in-in formalism,
 	\begin{align}
		\langle \varphi_{\bf k_1}(t)\varphi_{\bf k_2}(t)\varphi_{\bf k_3}(t) \rangle &\simeq i\int_0^t {\rm d}t'~\langle [H_{I}^{(3)}(t'), \varphi_{\bf k_1}(t)\varphi_{\bf k_2}(t)\varphi_{\bf k_3}(t)] \rangle, \label{eq:3ininformula}
	\end{align}
at the leading order in the $q$-parameters.
 To see the contributions from the resonance, it is sufficient to consider the linear gauge transformation (\ref{eq:gt}), $\zeta=H\varphi/\dot{\phi}$. 
Then, in terms of the dimensionless bispectrum (\ref{eq:dlB}), the correction can be estimated as,
	\begin{align}\label{eq:deltaB}
		\Delta {\cal B}_{\zeta} \simeq \frac{\bar{\epsilon}_V}{4}I_{\bf k_1k_2k_3}^{(3)}.
	\end{align}
Here,
	\begin{align}
		I_{\bf k_1k_2k_3}^{(3)} \equiv {\rm Re}\left[ \frac{m}{H}\int_0^z{\rm d}z'~C^{(3)}_{\bf k_1k_2k_3}e^{-iK\tau'}\right] + \text{(5 perms)} , \quad  (K \equiv k_1+k_2+k_3), \label{eq:I3}
	\end{align}
where the coefficient function $C^{(3)}_{\bf k_1k_2k_3}$ is defined as,
	\begin{align}
		C^{(3)}_{\bf k_1k_2k_3} \equiv C^{(3O)}_{\bf k_1 k_2 k_3} - i\left(\frac{k_2+k_3}{am}\right)C^{(3I)}_{\bf k_1 k_2 k_3} - \left(\frac{k_2}{am}\right)\left(\frac{k_3}{am}\right)C^{(3II)}_{\bf k_1 k_2 k_3}.
	\end{align}
We have also introduced the averaged slow-roll parameter $\bar{\epsilon}_V$ by,
	\begin{align}
		\bar{\epsilon}_V \equiv \sqrt{\epsilon_{V,\ast}\epsilon_{V,c}},
	\end{align}
where $\epsilon_{V,\ast}$ and $\epsilon_{V,c}$ are evaluated at the resonance and the horizon crossing, respectively.
In deriving Eq. (\ref{eq:I3}), we have used the subhorizon approximation for all mode functions, $u_{k_i}(t')~(i=1,2,3)$, assuming that they were in the subhorizon regime at the resonance, $k_i/a_{\ast}H >1$.
This approximation is not valid when we consider the squeezed limit, where one of the modes can be outside of the horizon when the oscillation is excited. 
We consider this limit separately in the last part of this subsection.

\begin{figure}[t]
		\centering
		\includegraphics[width=0.9\linewidth]{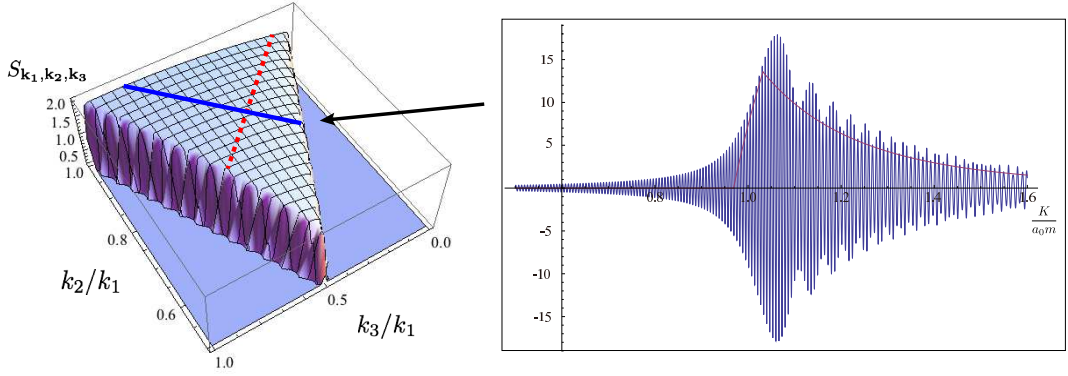}
		\caption{The resonant enhancement of the bispectrum through the couplings $K_n$. In the left panel, we showed the shape function (\ref{eq:shape}). Along the solid line, the bispectrum scales as shown in the right panel. The lines indicate the same in Fig. \ref{fig:power}. The bispectrum has large values along the dashed line, $K \equiv k_1+k_2+k_3=2k_p$, where $k_p$ is the peak scale of the feature in the power spectrum. Here, we have set $\Gamma=5H$, $m=10^3H$, $q_n=0.1$, and $\bar{\epsilon}_V=0.01$. If other interactions exist, larger enhancements can be realized (\S \ref{ss:SVI}).}
		\label{fig:bi}
	\end{figure}

Since the integral has a similar form as Eq. (\ref{eq:I2app}), we can estimate it as in the previous section. 
The oscillatory components, $e^{imt'}$ and $e^{2imt'}$, in $C_{\bf k_1k_2k_3}^{(3)}$ (see Eqs. (\ref{eq:C3koii}) and (\ref{eq:C3ki})) resonate with the factor from the mode functions, $e^{-iK\tau'}$.
The resonance occurs in a similar way as in the power spectrum, where the oscillatory components in the interactions resonate with the factor from the mode functions $e^{-2ik\tau'}$. 
Here, we evaluate $I_{\bf k_1k_2k_3}^{(3)}$ for the couplings $K_n$. 
Replacing the parameters $(q_n, m/2, \Gamma/2)$ by $(q_d,m,\Gamma)$, we can obtain a similar result for the couplings  $K_d$.
The phase function $\theta^{(3;n)} \equiv mt - K\tau$ has a stationary point at time $t_{\ast}$ where
	\begin{align}
		\frac{K}{a(t_{\ast})} = m.
	\end{align}
%for the $K_n$ couplings and
%	\begin{align}
%		\frac{K}{a(t_{\ast})} = 2m,
%	\end{align}
%for the $K_d$ couplings, respectively. 
Then, using the function (\ref{eq:Inorm}) with $K$ in the argument, the integral can be evaluated as,
	\begin{align}
		I_{\bf k_1k_2k_3}^{(3)} \simeq {\rm Re}I_K^{(n)}\left[C^{(3O)}_{\bf k_1 k_2 k_3,\ast} - \frac{k_2+k_3}{K}C^{(3I)}_{\bf k_1 k_2 k_3,\ast} - \left(\frac{k_2}{K}\right)\left(\frac{k_3}{K}\right)C^{(3II)}_{\bf k_1 k_2 k_3,\ast}\right]\frac{m}{H},
	\end{align}
where the coefficient functions are evaluated at the resonance, $t=t_{\ast}$.
The duration of the resonance is given by Eq. (\ref{eq:durationN}).
% and (\ref{eq:durationD}). 
Note that the bispectrum has contributions from the subhorizon regime \cite{Chen:2008wn, Flauger:2010ja, Chen:2010bka, Behbahani:2011it, Gwyn:2012pb} in contrast to the cases without a resonance.
Then, $I_{\bf k_1k_2k_3}^{(3)}$ can be approximated as,
	\begin{align}\label{eq:I3app}
		I_{\bf k_1k_2k_3}^{(3;n)} \simeq \cos\theta^{(3;n)}_{\ast}I_{K,\text{approx}}^{(n)}\left[C^{(3O)}_{\bf k_1 k_2 k_3,\ast} - \frac{k_2+k_3}{K}C^{(3I)}_{\bf k_1 k_2 k_3,\ast} - \left(\frac{k_2}{K}\right)\left(\frac{k_3}{K}\right)C^{(3II)}_{\bf k_1 k_2 k_3,\ast}\right]\frac{m}{H} \nonumber \\ + \text{(5 perms)} ,
	\end{align}
 and then the correction to the bispectrum (\ref{eq:deltaB}) can be roughly estimated as,
	\begin{align}\label{eq:correction3}
		\Delta {\cal B}_{\zeta}^{(n)} \simeq \frac{\bar{\epsilon}_V}{4}\left(\frac{m}{H}\right)q_n\cos\theta^{(3;n)}_{\ast} I_{K,\text{approx}}^{(n)}S_{\bf k_1,k_2,k_3},
	\end{align}
where
	\begin{align}\label{eq:shape}
		S_{\bf k_1,k_2,k_3} \equiv q_n^{-1}\left[C^{(3O)}_{\bf k_1 k_2 k_3,\ast} - \frac{k_2+k_3}{K}C^{(3I)}_{\bf k_1 k_2 k_3,\ast} - \left(\frac{k_2}{K}\right)\left(\frac{k_3}{K}\right)C^{(3II)}_{\bf k_1 k_2 k_3,\ast}\right] + \text{(5 perms)},
	\end{align}
for the couplings $K_n$. 
Therefore, the amplitude of the correction to the bispectrum can be written in terms of that to the power spectrum (\ref{eq:correction2zeta}) as,
	\begin{align}\label{eq:correction3order}
		\Delta {\cal B}_{\zeta}^{(n)} \sim \frac{\bar{\epsilon}_V}{4}\left(\frac{m}{H}\right)\frac{\Delta {\cal P}_{\zeta}^{(n)}}{{\cal P}_{\zeta}}S_{\bf k_1,k_2,k_3}.
	\end{align}
As is clear from the expression (\ref{eq:correction3order}), the correction to the bispectrum can be much larger than that to the power spectrum thanks to the large factor $m/H$.
%The contributions from the $K_d$ couplings can be obtained in a similar way. 
In Fig. \ref{fig:bi}, we showed the shape and scaling of the correction to the bispectrum. 
The bispectrum has large values for $K=2k_p$, where $k_p$ is the peak scale of the feature in the power spectrum, $k_p/a_0m \simeq e^{\sqrt{H/m}}/2$.
Therefore, the characteristic scale of the feature in the bispectrum is correlated with that in the power spectrum. 

%%%%%%%%%%%%%%%%%%%%%%%%%%%%%%%%%%%%%%
%%%%%%%%%%%%%%%%%%%%%%%%%%%%%%%%%%%%%%

To see the magnitude of the amplification, here, we compare the result (\ref{eq:correction3}) with the equilateral-form and local-form bispectra at the specific configurations. \\

- {\it Equilateral configurations} ($k_1 \simeq k_2 \simeq k_3 \simeq 2k_p/3$) \\

The amplitude of the correction is estimated to be 
	\begin{align}
		\Delta {\cal B}_{\zeta, \text{equilateral}}^{(n)} &\simeq \frac{7\bar{\epsilon}_{V}}{16}\left(\frac{m}{H}\right)\frac{\Delta {\cal P}_{\zeta}^{(n)}}{{\cal P}_{\zeta}} \\
		&\simeq \frac{7\bar{\epsilon}_{V}}{8}q_n\left(\frac{m}{H}\right)^{\frac{3}{2}}  \quad (\text{for } \Gamma \ll \sqrt{Hm}),
	\end{align}
at the equilateral point $k_1 \simeq k_2 \simeq k_3 \simeq 2k_p/3$, while for the equilateral-type $f_{\rm NL}$,
	\begin{align}
		{\cal B}_{\zeta, \text{equilateral}} = \frac{9}{10}f_{\rm NL}^{\rm equil}.
	\end{align} 
\\

- {\it Squeezed configurations} ($k_3 \ll k_1 \simeq k_2 \simeq k_p$)\\

In the finitely squeezed configurations at the specific scale, $k_3 \ll k_1 \simeq k_2 \simeq k_p$ but $k_3 > a_{\ast}H \simeq (H/m)K$, we obtain a non-vanishing contribution,
	\begin{align}
		\Delta {\cal B}_{\zeta, \text{{\it not so} squeezed}}^{(n)} &\simeq \frac{7\bar{\epsilon}_{V}}{16}\left(\frac{m}{H}\right)\frac{\Delta {\cal P}_{\zeta}^{(n)}}{{\cal P}_{\zeta}} \\
		&\simeq \frac{7\bar{\epsilon}_V}{8}q_{n}\left(\frac{m}{H}\right)^{\frac{3}{2}}  \quad (\text{for } \Gamma \ll \sqrt{Hm}),
	\end{align}
while for the local-type $f_{\rm NL}$,
	\begin{align}
		{\cal B}_{\zeta, \text{local}} = \frac{3}{10}f_{\rm NL}^{\rm local}\left(\frac{k_p}{k_3}\right).
	\end{align} 
 In terms of the bispectrum,
	\begin{align}
		&\langle \zeta_{{\bf k}_1}(t)\zeta_{{\bf k}_2}(t)\zeta_{{\bf k}_3}(t) \rangle_{ (2H/m)k_p \ll k_3 \ll k_p,~ k_1 \simeq k_2 \simeq k_p} \simeq \nonumber \\
		&\hspace*{8em} (2\pi)^3\delta^3\left({\bf k_1+k_2+k_3}\right)P_{\zeta}(k_1) \Delta P_{\zeta}^{(n)}(k_3)\left(\frac{k_3}{k_p}\right)\left[\frac{7\bar{\epsilon}_V}{16}\left(\frac{m}{H}\right)\right], \label{eq:notsosqueezed}
	\end{align}
where $P_{\zeta}$ is the power spectrum for $\zeta$, $P_{\zeta} \equiv 2\pi^2 {\cal P}_{\zeta}/k^3$. 
On the other hand, in the standard single-field slow-roll inflation models, the bispectrum for the squeezed configurations is of the order of the slow-roll parameters:
	\begin{align}\label{eq:consistencyrel}
		&\langle \zeta_{{\bf k}_1}(t)\zeta_{{\bf k}_2}(t)\zeta_{{\bf k}_3}(t) \rangle_{k_3 \ll k_1 \simeq k_2 \simeq k_p} = \nonumber \\
		&\hspace*{8em} -(2\pi)^3\delta^3\left({\bf k_1+k_2+k_3}\right)P_{\zeta}(k_1) P_{\zeta}(k_3)\left[ n_s -1 + {\cal O}\left(\frac{k_3}{k_p}\right)^2 \right],
	\end{align}
where $n_s$ is the spectral index of the power spectrum.
Hence, the excitation of a heavy scalar field can give a non-negligible effect on the bispectrum for the configurations, $(2H/m)k_p \ll k_3 \ll k_p,~ k_1 \simeq k_2 \simeq k_p$, as recently pointed out in Ref. \cite{Flauger:2013hra}. 
If the peak scale $k_p$ is not so large, the scales $k \ll (2H/m)k_p$ could be in the superhorizon regime at the present time.
In that case, these contributions could provide practically the most important one in observations of the bispectrum in the squeezed limit. 
Note that the contribution (\ref{eq:notsosqueezed}) corresponds to the ${\cal O}(k_3/k_p)^2$ correction in Eq. (\ref{eq:consistencyrel}).
This correction can be large in the resonant cases because it appears in the combination $(k_3\tau_{\ast})^2 \simeq (k_3/2k_p)^2(m/H)^2$, 
which is not necessarily small even for the squeezed configurations $k_3 \ll k_1 \simeq k_2$.

As mentioned in the text below Eq. (\ref{eq:I3}), our subhorizon approximation is not valid in the squeezed limit where $k_3 \ll (2H/m)k_p,~k_1 \simeq k_2\simeq k_p$.
In this case, the mode function $u_{k_3}$ and its time derivative $\dot{u}_{k_3}$ should be replaced by those with the extra factors $i/k\tau'$ and $1/(k\tau')^2$, respectively.
Hence, the contributions to the squeezed limit can be estimated as,
	\begin{align}
		\Delta {\cal B}_{\zeta, \text{squeezed}}^{(n)} &\sim \bar{\epsilon}_{V}\left(\frac{k_p}{k_3}\right)\frac{\Delta {\cal P}_{\zeta}^{(n)}}{{\cal P}_{\zeta}} \\
		&\sim \bar{\epsilon}_V q_{n}\left(\frac{k_p}{k_3}\right)\left(\frac{m}{H}\right)^{\frac{1}{2}}  \quad (\text{for } \Gamma \ll \sqrt{Hm}),
	\end{align}
and the bispectrum behaves as,
	\begin{align}
		\langle \zeta_{{\bf k}_1}(t)\zeta_{{\bf k}_2}(t)\zeta_{{\bf k}_3}(t) \rangle_{k_3 \ll k_1 \simeq k_2 \simeq k_p} \sim (2\pi)^3\delta^3\left({\bf k_1+k_2+k_3}\right)P_{\zeta}(k_1) \Delta P_{\zeta}^{(n)}(k_3)\epsilon_V.
	\end{align}
Therefore, the bispectrum scales as in the usual case (Eq. (\ref{eq:consistencyrel})) and cannot become large unless the modification in the power spectrum is large. \\

As discussed in the previous section, we need to require the conditions $\Delta {\cal P}_{\zeta}^{(n)}/{\cal P}_{\zeta}<1$ and $m < 2\pi \Lambda_n$ for the perturbativity. 
The correction has the maximum value when these conditions are saturated. 
In this case, both energy scales $m$ and $2\pi\Lambda_n$ are given by ${\cal O}(10^3)(f_{\chi,t=0}/\bar{\epsilon}_V)^{1/3}H$.
Then, our perturbative calculation is reliable if
	\begin{align}
		\Delta {\cal B}_{\zeta}^{(n)} \sim \bar{\epsilon}_V\left(\frac{m}{H}\right)\frac{\Delta {\cal P}_{\zeta}^{(n)}}{{\cal P}_{\zeta}} < {\cal O}(10^3) \times  (\bar{\epsilon}_V^2 f_{\chi,t=0})^{\frac{1}{3}}.
	\end{align}
Here, we have used the observed value for the amplitude of the power spectrum, ${\cal P}_{\zeta} = {\cal O}(10^{-9})$ \cite{Hinshaw:2012fq, Ade:2013uln}.
Similarly, in the case of the couplings $K_d$ , we can find the upper limit for the perturbativity as,
	\begin{align}
		\Delta {\cal B}_{\zeta}^{(d)} \sim \bar{\epsilon}_V\left(\frac{m}{H}\right)\frac{\Delta {\cal P}_{\zeta}^{(d)}}{{\cal P}_{\zeta}} < {\cal O}(10^3) \times (\bar{\epsilon}_V^5 f_{\chi,t=0}^2)^{\frac{1}{7}}.
	\end{align}
The energy scales $m$ and $(2\pi)^{1/2}\Lambda_d$ are given by ${\cal O}(10^3)(f_{\chi,t=0}/\bar{\epsilon}_V)^{2/7}H$ when the inequality is saturated.

Therefore, the correction can be ${\cal O}(10^{1-2})$, which is much larger than that expected in the standard single-field slow-roll inflation models, 
 even when the correction to the power spectrum is small. 
More larger non-Gaussianity can be induced when the correction to the power spectrum is measurably large.

%%%%%%%%%%%%%%%%%%%%%%%%%%%%%%%%
	\subsection{Perturbativity}

As in the case of the power spectrum, the higher-order terms could be comparable to the leading term even when the $q$-parameters are small. 
Here, we discuss the validity of the perturbation for the cubic interactions (\ref{eq:intHphi}).

  First, we estimate the magnitude of the interaction Hamiltonian (\ref{eq:intHphi}) in the resonance regime from the dimensional argument. 
  The interaction Hamiltonian (\ref{eq:intHphi}) can be estimated as,
  	\begin{align}\label{eq:intH3mag}
		H_I^{(3)} \sim \frac{aq\sqrt{\epsilon_V}}{(2\pi)^6M_p}k^8\varphi_{\bf k}^3,
	\end{align}
where we have assumed that the wavenumbers have the same order $k$.
The fluctuations $\varphi_{\bf k}$ can be estimated as,
	\begin{align}\label{eq:fmag}
		\varphi_{\bf k} &\sim \frac{1}{a\sqrt{k}}\left(\frac{2\pi}{k}\right)^{\frac{3}{2}}.
	\end{align}
Here, we have replaced the creation/annihilation operators by $(2\pi/k)^{\frac{3}{2}}$, taking into account their commutation relation $[a_{\bf k},{a_{\bf k'}}^{\dagger}] = (2\pi)^3\delta^3({\bf k-k'})$.
As mentioned in the previous section, the order parameter of the expansion is $H_I^{(3)}\Delta t$, where $\Delta t \sim 1/\sqrt{mH}$. 
Using Eqs. (\ref{eq:intH3mag}) and (\ref{eq:fmag}), it can be estimated as,
	\begin{align}
		H_I^{(3)}\Delta t &\sim (2\pi)^{-\frac{3}{2}}q\sqrt{\epsilon_V}\left(\frac{H}{M_p}\right)\left(\frac{m}{H}\right)^{\frac{3}{2}} \\
		&\sim \left(\frac{{\cal P}_{\zeta}}{2\pi}\right)^{\frac{1}{2}}\Delta {\cal B}_{\zeta},
	\end{align}
at the resonance, $k/a \sim m$.
Therefore, the higher-order terms are subdominant for the cubic interactions as long as $\Delta {\cal B}_{\zeta} < {\cal O}(10^5)$. 

The quadratic interaction Hamiltonian (\ref{eq:intH2}) also gives corrections to Eq. (\ref{eq:3ininformula}). 
In a similar way as above, the order parameter of the corrections is estimated to be,
	\begin{align}
		H_I^{(2)}\Delta t \sim q\sqrt{\frac{m}{H}}.
	\end{align}
Hence, we can use the perturbation safely when the feature in the power spectrum is negligibly small, $\Delta {\cal P}_{\zeta}/{\cal P}_{\zeta} \sim q\sqrt{m/H} < 1$, as discussed in the previous section. 
When this condition is not satisfied, we should solve the full equation of motion to get the mode function. 
In that case, the mode function contains the negative energy mode and the bispectrum has additional resonant contributions \cite{Chen:2010bka}. 
We can also estimate the cutoff scale $\Lambda_{\rm cut}$ discussed in the last part of \S\ref{ss:power} by applying a similar argument to the derivative couplings with $\Delta t =2\pi/m$.

%%%%%%%%%%%%%%%%%%%%%%%%%%%%%%%%%%	
	\subsection{Other Interactions}\label{ss:SVI}
%%%%%%%%%%%%%%%%%%%%%%%%%%%%%%%%%%
 In the previous subsections, we have seen that the bispectrum can be affected much by the resonance through the couplings $K_n$ and $K_d$ at the specific scale.
 However, as can be seen in Eq. (\ref{eq:intHphi}), their effects are suppressed by the Planck scale.
 This is because cubic interactions of the inflaton field are not contained in the couplings $K_n$ and $K_d$. 
 Then, they are induced through the gravitational couplings. 
 Direct cubic interactions appear if there is the self interaction of the inflaton field,
 	\begin{align}\label{eq:self}
		\frac{\lambda_s}{4\Lambda_d^4}(\pd \phi)^4.
	\end{align}
 They are also induced by the interactions that are higher order in $\Lambda_d$,
 	\begin{align}\label{eq:higher}
		\frac{\lambda_h}{8\Lambda_d^8}(\pd \chi \cdot \pd \phi)^2(\pd \phi)^2,
	\end{align}
or violate the parity symmetry $\phi \to -\phi$,
	\begin{align}\label{eq:parity_violating}
		\frac{\lambda_p}{4\Lambda_d^4}(\pd \chi \cdot \pd \phi)(\pd \phi)^2.
	\end{align}
 Here, we estimate the magnitude of their effects on the bispectrum. 
  The effects of the interactions (\ref{eq:self}), (\ref{eq:higher}), and (\ref{eq:parity_violating}) on the background evolution and the power spectrum can be analyzed in a similar way as the couplings $K_d$; 
 The interactions (\ref{eq:self}) and  (\ref{eq:higher}) have little effects because they are higher order in the $q$-parameters, 
 while the interaction (\ref{eq:parity_violating}) could have an effect as large as the couplings  $K_d$ with the $q$-parameter of the order of $q(\epsilon_V f_{\chi,t=0}^{-1})^{\frac{1}{2}}$. 
 In the parity-vaiolating case, the kinetic mixing term $(\pd \chi \cdot \pd \phi)$ could appear. 
 We assume that this term is so suppressed that the background evolution and the power spectrum are not affected much. 
 
 The interaction (\ref{eq:self}) induce the interaction Hamiltonian as,
 	\begin{align}
		a^3\left(\frac{\lambda_s \dot{\phi}}{\Lambda_d^4}\right)\int {\rm d}^3x~ \dot{\varphi}(\pd \varphi)^2.
	\end{align}
Oscillatory components are induced in this Hamiltonian at the higher order in the $q$-parameters (see Eqs. (\ref{eq:ibgsol}) and (\ref{eq:icm})).

 On the other hand, the interactions (\ref{eq:higher}) and (\ref{eq:parity_violating}) induce the interaction Hamiltonian as,
 	\begin{align}
		a^3\left(\frac{\lambda_h \dot{\chi}^2\dot{\phi}}{4\Lambda_d^8}\right)\int {\rm d}^3x~ \dot{\varphi}\left[2\dot{\varphi}^2 - \frac{1}{a^2}(\nabla \varphi)^2 \right],
	\end{align}
 and
 	\begin{align}
		a^3\left(\frac{\lambda_p \dot{\chi}}{4\Lambda_d^4}\right)\int {\rm d}^3x~ \dot{\varphi}(\pd \varphi)^2,
	\end{align}
respectively. 
Using the dimensional argument in the previous subsection, their effects on the bispectrum can be estimated to be,
	\begin{align}
		\Delta {\cal B}^{(s)}_{\zeta} \sim \Delta {\cal B}^{(h)}_{\zeta} &\sim q^2(\epsilon_V f_{\chi,t=0}^{-1})\left(\frac{m}{H}\right)^{\frac{5}{2}} \sim (\epsilon_V f_{\chi,t=0}^{-1})\left(\frac{m}{H}\right)^{\frac{3}{2}}\left(\frac{\Delta {\cal P}_{\zeta}}{{\cal P}_{\zeta}}\right)^2,\\
		\Delta {\cal B}^{(p)}_{\zeta} &\sim q(\epsilon_V f_{\chi,t=0}^{-1})^{\frac{1}{2}}\left(\frac{m}{H}\right)^{\frac{5}{2}} \sim (\epsilon_V f_{\chi,t=0}^{-1})^{\frac{1}{2}}\left(\frac{m}{H}\right)^{2}\frac{\Delta {\cal P}_{\zeta}}{{\cal P}_{\zeta}},
	\end{align}
at the peak scale. 
Therefore, the corrections to the bispectrum through the interactions (\ref{eq:self}), (\ref{eq:higher}), and (\ref{eq:parity_violating}) are enhanced respectively by the factors $(m/H)^{3/2}$ and $(m/H)^2$, which can be much larger than the factor $m/H$ for the couplings $K_n$ and $K_d$.
Since all mode functions appear with a derivative, these contributions scale as $k_3^{-1}$ for finitely squeezed configurations where the subhorizon approximation (\ref{eq:MFsub}) is appropriate. 
In the squeezed limit, on the other hand, we cannot use the subhorizon approximation and the bispectrum scales as $k_3^{-3}$ as in the usual cases.
	
If we require the conditions $\Delta {\cal P}_{\zeta}^{(d)}/{\cal P}_{\zeta}<1$ and $m < (2\pi)^{1/2} \Lambda_d$ for the perturbativity, 
they should be bounded from above as,
	\begin{align}
		\Delta {\cal B}^{(s)}_{\zeta}, ~ \Delta {\cal B}^{(h)}_{\zeta} &< {\cal O}(10^5) \times (\epsilon_V f_{\chi,t=0}^{-1})^{\frac{4}{7}}, \\
		\Delta {\cal B}^{(p)}_{\zeta} &<  {\cal O}(10^6) \times (\epsilon_V f_{\chi,t=0}^{-1})^{-\frac{1}{14}}.
	\end{align}
Therefore, the interactions (\ref{eq:higher}) and (\ref{eq:parity_violating}) could induce large features in the bispectrum even when the features in the power spectrum are too small to be detected. 
Note that the scale of the resonance induced by the interaction (\ref{eq:parity_violating}) is different from that for the couplings $K_d$;
The resonance occurs at $K/a \sim m$ for the interaction (\ref{eq:parity_violating}) while $K/a \sim 2m$ for the couplings $K_d$.
If we consider much higher-order interactions, the scales of the resonance appear at the integral multiples of the mass scale.
Though they are suppressed by the $q$-parameters, they could still induce non-negligible features in the bispectrum.

%%%%%%%%%%%%%%%%%%%%%%%%%%%%%%%%%
\section{Summary and Discussion}\label{s:summary}
%%%%%%%%%%%%%%%%%%%%%%%%%%%%%%%%%
Non-Gaussianity could contain various information on the physics behind inflation. 
In this paper, we have discussed the possibility that we could obtain hints on the heavy physics during inflation by analyzing local features in the primordial bispectrum. 
A heavy scalar field can leave non-negligible signatures in the primordial spectra through the parametric resonance between its background oscillation and the fluctuations in the inflaton field. 
We have estimated the contributions from the resonance perturbatively by picking up the interactions relevant to it. 
The bispectrum is amplified at specific configurations, and 
its amplitude can be ${\cal O}(10^{1-2})$, or as large as ${\cal O}(10^5)$ depending on the type of interactions within the parameter region where the perturbative expansion is applicable.
In particular, the resonance can give large contributions in finitely squeezed configurations, while the bispectrum cannot be large in the squeezed limit unless the modification to the power spectrum is large as in the case that the consistency relation is satisfied. 
Since we cannot observe the exact squeezed limit, these contributions could practically be most important in actual observations of the bispectrum in that limit. 
In the analysis, we have assumed that oscillations of a heavy scalar field are excited without introducing other effects.
This assumption may be too idealistic when we consider a specific excitation mechanism of the oscillations.
In general, there might be slow-roll violations \cite{Avgoustidis:2012yc}, excitations from the vacuum, or mixing of the light and heavy fields through non-derivative couplings \cite{Cremonini:2010ua, Achucarro:2010jv, Achucarro:2010da, Cespedes:2012hu, Achucarro:2012sm, Avgoustidis:2012yc, Chen:2012ge, Pi:2012gf, Burgess:2012dz, Noumi:2012vr, Shiu:2011qw, Gao:2012}.
We will explore their effects on the resonance in future work.

 We have also found that there is a relation between the scales at which the features appear in the bispectrum and the power spectrum, and that the feature in the bispectrum can be much larger than that in the power spectrum. 
 Moreover, if we consider a specific excitation mechanism like a turn in the inflaton trajectory, other features could be also induced in the primordial spectra around the horizon scale at the excitation time. In particular, as discussed in Ref. \cite{Achucarro:2012fd}, correlated features in the bispectrum and the power spectrum are induced by a turn in the inflaton trajectory.
 Since the resonance scale is comparable to the mass scale at the excitation time, we could determine the mass of the heavy scalar field if both of these features are detected.
Though we have not discussed the observability of localized features in the bispectrum,  it will be discussed in the upcoming paper by one of the present author \cite{YTJY}. 
If the correlated features are observed at the characteristic scales in the primordial spectra, it will indicate the presence of heavy degrees of freedom. 
 By analyzing them, we could obtain some information on the physics behind inflation.

\section*{Acknowledgement}
 We thank J. Yokoyama for initial collaboration and J. R. White for reading a part of the manuscript.
 The work is supported by a Grant-in-Aid through JSPS (No. 23-3430 and No. 24-2236).
 
\appendix
\section{Interactions Relevant to the Resonance}\label{s:RI}
Here, we discuss which cubic interactions are most relevant to the resonance. 
The cubic interactions for the action (\ref{eq:generalaction}) have been obtained in Refs. \cite{Langlois:2008qf, Arroja:2008yy}.
They can be separated into the gravitational and matter parts as,
	\begin{align}
		H_g^{(3)} &=  -a^3M_p^2H^2\int{\rm d}^3x~\left[3\alpha^2 + \frac{2}{a^2H}\alpha\nabla^2\beta + \frac{1}{a^4H^2}(\nabla^2\beta\nabla^2\beta - \nabla_i\nabla_j\beta\nabla^i\nabla^j\beta) \right]\alpha,\\
		H_m^{(3)} &= -a^3\int{\rm d}^3x~\left[ P_{IJ}(X_3^{IJ} + \alpha X_2^{IJ}) + \frac{1}{2}P_{IJ,KL}(2X_2^{IJ} + \alpha X_1^{IJ} )X_1^{KL} \right. \nonumber \\
		&\quad + P_{IJ,K}(X_2^{IJ} + \alpha X_1^{IJ})\varphi^K + \frac{1}{2}P_{I,J}\alpha \varphi^I\varphi^J + \frac{1}{6}P_{IJ,KL,MN}X_1^{IJ}X_1^{KL}X_1^{MN} \nonumber \\
		&\quad \left. + \frac{1}{2}P_{IJ,KL,M}X_1^{IJ}X_1^{KL}\varphi^{M} + \frac{1}{2}P_{IJ,K,L}X_1^{IJ}\varphi^{K}\varphi^{L} + \frac{1}{6}P_{I,J,K}\varphi^I\varphi^J\varphi^K \right].
	\end{align}
Here, $\alpha$ and $\beta$ are respectively the perturbations in the lapse and shift functions at the first order, which are given by solving the constraint equations as,
	\begin{align}
		\alpha &= \frac{\pi_I}{2M_p^2H}\varphi^I, \\
		\frac{\nabla^2\beta}{a^2 H} &= -3\alpha + \frac{1}{2M_p^2H^2}\left[P_I\varphi^I - P_{IJ,K}\dot{\phi}^I\dot{\phi}^J\varphi^K + (P_{IJ}+P_{IJ,LM}\dot{\phi}^L\dot{\phi}^M)(\dot{\phi}^I\dot{\phi}^J\alpha-\dot{\phi}^I\dot{\varphi}^J) \right].
	\end{align}
The function $X_n^{IJ}~(n=1,2,3)$ represents the perturbations in $X^{IJ}\equiv-(\pd_{\mu} \phi^I)(\pd^{\mu}\phi^J)/2$ at the $n$-th order. They can be read from
	\begin{align}
		X^{IJ} &= \frac{1}{2(1+\alpha)^2}\sum_{n=0}^4 V_n^{IJ} - \frac{1}{2a^2}\nabla_i\varphi^I\nabla^i\varphi^J,
	\end{align}
where
	\begin{align}
		V_0^{IJ} &= \dot{\phi}^I\dot{\phi}^J, \\
		V_1^{IJ} &= 2\dot{\phi}^{(I}\dot{\varphi}^{J)}, \\
		V_2^{IJ} &= \dot{\varphi}^I\dot{\varphi}^J - \frac{2}{a^2}\nabla^i\beta\dot{\phi}^{(I}\nabla_i\varphi^{J)}, \\
		V_3^{IJ} &= -\frac{2}{a^2}\nabla^i\beta\dot{\varphi}^{(I}\nabla_i\varphi^{J)},\\
		V_4^{IJ} &= \frac{1}{a^4}(\nabla_i\beta\nabla^i\varphi^I)(\nabla_i\beta\nabla^i\varphi^J).
	\end{align}
Here, $()$ in the superscript represents the symmetrization.

 Since the resonance occurs deep in the horizon, the interactions with more derivatives are more relevant to the resonant enhancement of the bispectrum. 
 Taking the terms with more derivatives, the lapse and shift functions are estimated to be,
 	\begin{align}
		\alpha &\simeq -\sqrt{\frac{\epsilon_V}{2}}\frac{\varphi}{M_p}, \\
		\frac{\nabla^2\beta}{a^2 H} &\simeq \sqrt{\frac{\epsilon_V}{2}}\frac{\dot{\varphi}}{M_pH}\frac{\hat{z}_{\phi}^2 + 2(q_{d1}+q_{d2})e^{-2\Gamma t}\sin^2(mt)}{\hat{z}_{\phi}^2},
	\end{align}
at the leading order in the slow-roll parameter $\epsilon_V$ and the fraction $f_{\chi}$. 
Though the energy fraction of the heavy scalar field $f_{\chi}$ contains an oscillatory component, 
\footnote{In the case that $\Gamma \ll H$, the suppression factor should be replaced by $H/m$.}
	\begin{align}
		f_{\chi} \supset \frac{1}{2}\left(\frac{\Gamma}{m}\right)f_{\chi,t=0}e^{-2\Gamma t}\sin(2mt) ,
	\end{align}
we have neglected it because the decay rate $\Gamma$ is assumed to be much smaller than the mass scale $m$,
 and then it cannot lead to a large enhancement.

Counting the number of derivatives on $\varphi$, we can find the $(\pd \varphi)^3$-type interactions in the terms
	\begin{align}
		\frac{1}{6}P_{IJ,KL,MN}X_1^{IJ}X_1^{KL}X_1^{MN}, \quad P_{IJ,KL}X_2^{IJ}X_1^{KL}.
	\end{align}
However, these terms are absent unless we include the interaction $(\pd \phi)^4$, or those that are higher order in $\Lambda_d$ or violate the parity symmetry $\phi \to -\phi$.
In \S\ref{ss:SVI}, we discuss the resonant enhancement of the bispectrum induced by the $(\pd \varphi)^3$-type interactions.

The next candidates are provided by the $\varphi(\pd \varphi)^2$-type interactions. 
They can be found in the terms
	\begin{align}
		\frac{1}{a^4H^2}(\nabla^2\beta\nabla^2\beta - \nabla_i\nabla_j\beta\nabla^i\nabla^j\beta)\alpha, \quad  P_{IJ,K}X_2^{IJ}\varphi^K, \label{eq:twod1} \\
		\frac{1}{2}P_{IJ,KL}(2X_2^{IJ} + \alpha X_1^{IJ} )X_1^{KL}, \quad P_{IJ}(X_3^{IJ} + \alpha X_2^{IJ}). \label{eq:twod2}
	\end{align}
The first term is higher order in the slow-roll parameter $\epsilon_V$ and the second term is absent unless parity-violating interactions like $\chi\phi(\pd \phi)^2$ are introduced. 
Hence, the last two terms (\ref{eq:twod2}) provide the cubic interactions that are most relevant to the resonance when only the couplings $K_n$ and $K_d$ exist. 
In terms of $z_{\phi}$ and $c_s$, these terms can be explicitly written as,
	\begin{align}
		H_{\rm relevant}^{(3)} &= z_{\phi}^2\int {\rm d}^3x~\left\{ \frac{\alpha}{2}\left[  (3-2c_s^2)\dot{\varphi}^2 + \frac{c_s^2}{a^2}(\nabla \varphi)^2 \right] + \frac{1}{a^2}\nabla^i\beta\nabla_i\varphi\dot{\varphi}  \right\} \\
		&\simeq \frac{a^3}{M_p}\sqrt{\frac{\epsilon_V}{2}}\int {\rm d}^3x~\biggl\{ -\frac{\hat{z}_{\phi}^2}{2}\left[  (3-2c_s^2)\dot{\varphi}^2 + \frac{c_s^2}{a^2}(\nabla \varphi)^2 \right]\varphi \nonumber \\
		&\qquad +  \left[\hat{z}_{\phi}^2 + 2(q_{d1}+q_{d2})e^{-2\Gamma t}\sin^2(mt)\right]\nabla^i(\nabla^{-2}\dot\varphi)\nabla_i\varphi\dot{\varphi}  \biggr\}.
	\end{align}

\end{document}